# Study on the deconfined degree of freedom $g_1$ and the running coupling constant $\alpha_s(T)$

**A. G. Shalaby**

**Department of Physics, Faculty of Science, Benha University, Benha, Egypt.**

## Abstract

The degree of freedom in the confined hadronic matter phase and the deconifned phase i.e the quark gluon plasma (QGP) is important in the study of phase transition in the early universe. It is calculated according to the strong coupling constant. But in the present work we try to figure out the effect of the running coupling constant in the calculation of the degree of freedom in the confined-deconfined phase of matter.

**Keywords:** Degree of freedom, confined phase (hadronic matter), deconfined phase (QGP), strong coupling constant, running coupling constant.

## 1. Introduction

The strong coupling constant, $\alpha_s$, is one of the fundamental parameters of the Standard Model of Particle Physics. The energy dependence of $\alpha_s$ is predicted by the renormalization group equation (RGE). The value of $\alpha_s$ has been determined in many different processes, including a large number of results from hadronic jet production, in either $e^+e^-$ annihilation or in deep-inelastic *ep* scattering (DIS) up to energies of 209 ~ GeV [1]. In QCD, one has a single coupling constant $g_s$, or the usually more convenient $\alpha_s = \frac{g_s^2}{4\pi}$ , and various quark masses $m_f$ with f = u, d, ..., t. One refers to their dependence on $\mu$ in the framework of a given renormalization scheme (RS) ($\alpha_s(\mu^2)$, $m_f(\mu^2)$,... ) as to the running coupling constant to the running masses and so on [2]. The 'freezing' of quark–gluon 'color' deconfined degrees of freedom is the essential ingredient in determining the conditions in a transition between phases that has time to develop into equilibrium. The following discussion tacitly assumes the presence of latent heat *B* in the transition, and a discontinuity in the number of degrees of freedom, $g_2 = g_1$, where '1' refers to the primeval QGP phase and '2' to the final hadronic-gas state [3].

**Email :-** **asmaa.shalaby@fsc.bu.edu.eg; asmaa.gaber76@gmail.com**

The polarization of QCD vacuum causes a variation of the physical coupling under changes of distance ~ 1/*Q*, so QCD predicts a dependence $\alpha_S = \frac{g^2}{4\pi} = \alpha_S(Q)$.

This dependence is described theoretically by the renormalization group equations and determined experimentally at relatively high energies. However, the well-established conventional perturbation theory cannot be used effectively in the IR domain. Meanwhile, there exists a phenomenological indication in favor of a smooth transition from short distance to long distance physics [4].

## 2. QCD properties

### i Confinement

The color singlet states exist only in QCD vacuum (hadronic world), and represented by quark-antiquark bound states (mesons) and three-quarks (antiquarks) bound states (baryons and antibaryons) [5]. This suggests that the interactions between quarks and gluons must be strong at large distance scales and the potential between quarks increases with the quark separation.

It is shown in [6] that, the electric field generated by the two opposite charges which are in the (normal) vacuum; fig. (1a); and in the dielectric medium fig. (1b). Using the Gauss theorem one immediately finds the electric field as; $E = q/\sigma$, where q denotes the charge and $\sigma$ is the cross section of the tube. If $\sigma$ is independent of the distance r between the charges, their potential energy equals

$$V(r) = \frac{q^2}{\sigma} r \qquad \textbf{(1)}$$

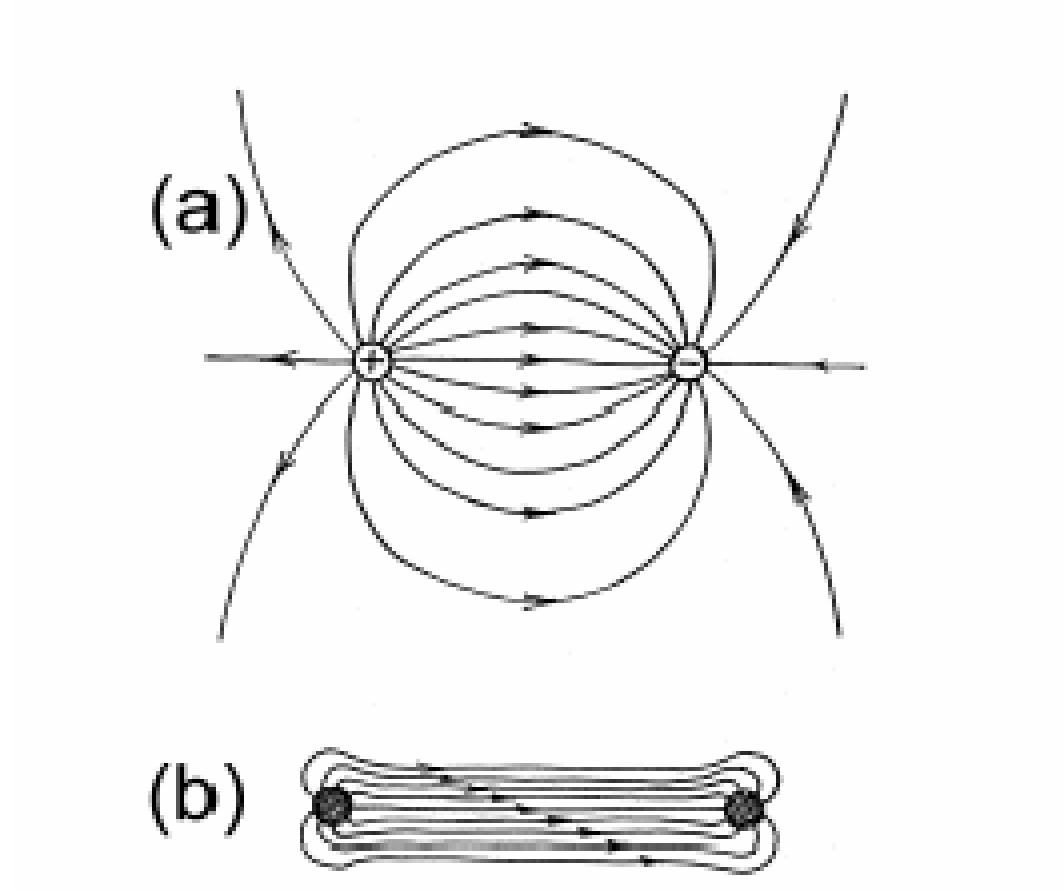

**Fig. (1) The electric field lines in the vacuum (a) and in dielectric medium (b)**

From eq. (1) one can see that, the potential energy grows linearly with r, when the charges are put in the dielectric medium.

Let us imagine that one tries to burst the meson up separating the quark from the antiquark. Stretching the meson requires pumping of the energy to the system. When the energy is sufficient to produce the quark–antiquark pair, the string breaks down and we have two mesons instead of one [6].

**ii <u>Asymptotic Freedom</u>**

As the quarks within a meson or baryon get closer together, the force of confinement gets weaker so that it asymptotically approaches zero for close confinement. The implication is that the quarks in close confinement are completely free to move around. Asymptotic freedom states that the coupling constant, which characterizes the strength of the quark gluon interaction in QCD, becomes weak at large relative momenta or at short distances [7]. The interaction between particles mediated by the gauge fields vanishes as the distance between the particles tends to zero (or the square of the four-momentum transfer between the particles tends to infinity). Since the particles behave as free particles in the asymptotically high energy region, this behavior is called asymptotic freedom. This feature provides a natural explanation for the parton model of hadrons [7].

Part of the nature of quark confinement is that the further you try to force the quarks apart, the greater the force of containment. A potential function which has been successfully used to describe some quark systems is of the form:

$$V = \frac{-K_1}{r} + K_2\, r \qquad \textbf{(2)}$$

Where, $K_1$ is the strength of Coulomb-like attraction of the quarks and $K_2$ the strength of the color force interaction.

The quark - quark coupling strength decreases for small values of r, as a resulting from the penetration of the gluon cloud surrounding the quarks [8]. The gluons carry "color charge" and therefore the penetration of the cloud would reduce the effective color charge of the quark.

## 2. The phase transition point

To find the phase transition point, one determines the (critical) temperature at which the pressures in the two phases are equal. One allows, in a transition of first order, for a difference in energy density $\varepsilon_1 \neq \varepsilon_2$ associated with the appearance of latent heat *B* (the 'bag constant'), which also enters the pressure of the deconfined phase [3]. One considers the Stefan Boltzmann pressure of a massless photon-like gas with degeneracy.

$$P_c \equiv P_1(T_c) = \frac{\pi^2}{90} g_1\, T_c^2 - B \qquad \textbf{(3)}$$

$$P_c \equiv P_2(T_c) = \frac{\pi^2}{90} g_1\, T_c^4 \qquad \textbf{(4)}$$

Then, one obtains the latent heat:

$$\frac{B}{T_c^4} = \frac{\pi^2}{90} \Delta g \qquad \textbf{(5)}$$

$$T_c = B^{1/4}\left(\frac{\pi^2}{90}\Delta g\right)^{1/4} \quad , \Delta g = g_1 - g_2 \qquad \textbf{(6)}$$

For the pressure at the transition temperature $T_c$, then one can determine the critical pressure $P_c$

$$P_c = B\frac{g_2}{\Delta g} \qquad \textbf{(7)}$$

The pressure, and therefore the dynamics of the transition in the early Universe, depends on the presence of non-hadronic degrees of freedom, which are absent from laboratory experiments with heavy ions. In summary, the phase-transition dynamics in the early Universe is determined by

(a) the effective number of confined degrees of freedom, $g_2$ at $T_c$.

(b) the change in the number of acting degrees of freedom $\Delta g$, which occurs exclusively in the strong-interaction sector.

(c) the vacuum pressure (latent heat) *B*, a property of strong interactions.

Both phases involved in the hadronization transition contain effectively massless electro-weak (EW) particles. Even though the critical temperature does not depend on the background of EW particles not participating in the transition, the value of the critical pressure, Eq. (7) depends on this, and thus to consider the active electro-weak degrees of freedom. These involve photons, $\gamma$, and all light fermions, viz., $e$, μ, $\nu_e$, $\nu_\mu$, and $\nu_\tau$ (one excludes the heavy $\tau$ -lepton with $m_\tau >> T$, and one considers the muon as being effectively a massless particle). Near to $T \approx$ 200 MeV, one obtains

$$g^{EW} = g_\gamma + \frac{7}{4} g_F^{EW} \; , \; g_\gamma = 2 \qquad \textbf{(8)}$$

And,

$$g_F^{EW} = \frac{7}{8} \times 2 \times \left(2_e + 2_\mu + 3_\nu\right) = 12.25 \qquad \textbf{(9)}$$

Where charged, effectively massless fermions enter with spin multiplicity 2, and one has three neutrino flavors – there are only left-handed light neutrinos and right-handed antineutrinos, and thus only half as many neutrino degrees of freedom as would naively be expected. In the deconfined QGP phase of the early Universe, one has

$$g_1 = g^{EW} + g_g + \frac{7}{4} g_q \qquad \textbf{(10)}$$

The number of effectively present strongly interacting degrees of freedom of quarks and gluons is influenced by their interactions, characterized by the strong coupling constant as,

$$g_g = 2_s \times 8_c \left(1 - \frac{15}{4\pi}\alpha_s\right) \qquad \textbf{(11)}$$

Where, $\frac{7}{4} g_q = \frac{7}{4}\, 2_s \times 2.5_f \times 3_c \left(1 - \frac{50}{21\pi}\alpha_s\right)$

where the flavor degeneracy factor used is 2.5, allowing in a qualitative manner for the contribution of more massive strangeness.

The degeneracies of quarks and gluons are indicated by the subscripts s (spin) and, c (color), respectively. one obtains for constant values of the strong coupling constant,

$$g_1 = \begin{cases} 56.5 & \text{for } \alpha_s = 0 \\ \approx 37 & \text{for } \alpha_s = 0.5 \\ \approx 33 & \text{for } \alpha_s = 0.6 \end{cases} \qquad \textbf{(12)}$$

In the present work, we have assumed the running coupling constant $\alpha_s(T)$ which is a function depends on the temperature T, instead of the strong coupling $\alpha_s$. The running coupling constant is taken as [9],

$$\alpha_s(T) = \frac{2\pi}{(11 - \frac{2}{3} n_f)\ \ell n(\frac{T}{\Lambda_\sigma})} \qquad \textbf{(13)}$$

Where, $n_f$ is the number of quark flavors, ($n_f$ = 0, 2, 3). From lattice QCD computations [9, 10], the parameters $\Lambda_\sigma = \beta T_c$ where, $\beta$ = 0.104 ± 0.009 and the critical temperature; $T_c$; lies in the range (150 − 300) MeV. In the present work, $T_c$ is taken as, $T_c$= 200 MeV.

## 3. The quark-gluon plasma (QGP)

One of the main activities in high-energy and nuclear physics is the search for the so-called quark-gluon plasma, a new state of matter which should have existed a few microseconds after the Big Bang.

After the Big Bang, the universe was very hot and quarks were deconfined. As the Universe cooled the quark-gluon plasma disappeared in what may have been a decofinement-confinment transition [11].

One of the important questions in perturbative QCD at finite temperature *T* is what is the temperature dependence of the strong coupling constant, $\alpha_s(T)$ in which an early study has been done by Collins and Perry [12].

In the present work the well-known one-loop expression for the running coupling constant has been used eq (13). This form is simpler than the two-loops expression and contains mainly the same physical information that $\alpha_s(T)$ decreases logarithmically as *T* increases. More rigorous treatments based on the Renormalization Group Equations (RGE) were developed, with applications to perturbative QCD. Some of the calculations did not result in the logarithmical dependence of the coupling in the temperature [13].

In general, the definition of a running coupling in QCD is not unique beyond the validity range of 2-loop perturbation theory. This is quite apparent when defining the coupling in QCD or in terms of the free energy potential at ( $T = 0$) [14].

## 4. Results and Discussion

In the present work, we have considered the running coupling constant $\alpha_S(T)$ instead of the strong coupling constant $\alpha_S$ to calculate $g_1$, B, $\Delta g$ and $P_c$.

In fig. (1) firstly we have calculated the selected form of the running coupling constant $\alpha_S(T)$ versus temperature T /$T_c$ at different selected critical temperatures $T_c$= 150, 200 MeV and at number of quark flavor $n_f$ = 2 . The solid curve represents the calculation of the running coupling constant $\alpha_S(T)$ eq. (13) at $T_c$= 200 MeV, and the dashed curve is the same plotting at $T_c$ =150 MeV. One can see that the running coupling decreases logarithmically as *T* increases for different critical temperatures as well.

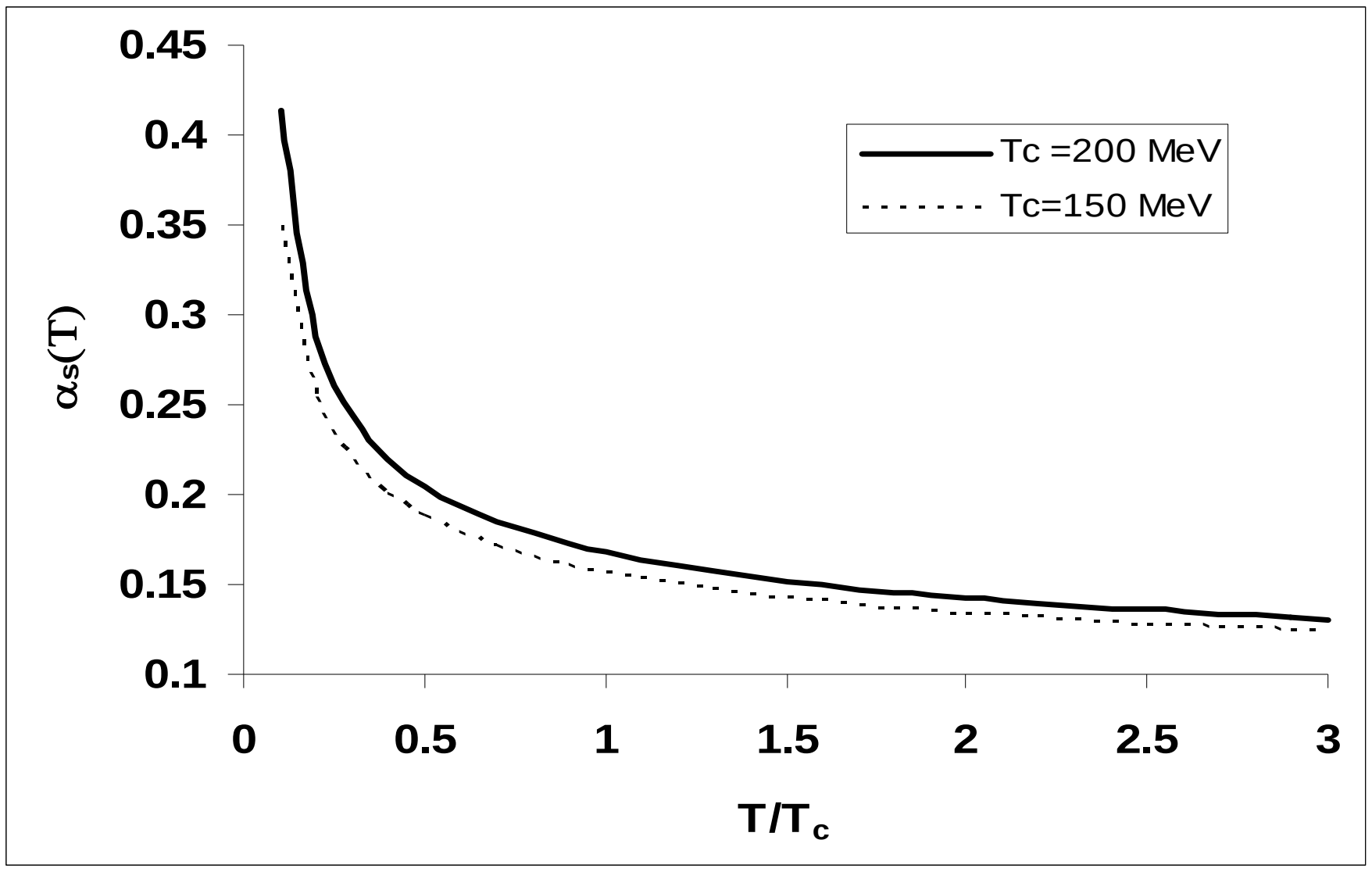

**Fig. (1) The running coupling constant $\alpha_S(T)$ versus T/$T_c$ and at different critical temperature $T_c$= 150, 200 MeV, $n_f$=2.**

Fig. (2) shows the behavior of the degree of freedom $g_1$ in the deconfined QGP phase [10], within the temperature normalized to the critical one $T_c$= 150, 200 MeV. Where '1' refers to the primeval QGP phase.. In this work we have inserted the running coupling constant $\alpha_S(T)$, which is a function of the temperature instead of a constant value of the strong coupling constant $\alpha_S$ .

It is obvious that the effect of the temperature in this formula according to the selected form of the running coupling constant eq. (13). The change appears in $g_g$ and $g_q$ calculations because of their dependence on $\alpha_s$. From fig. (2), It is obvious that the degree of freedom $g_1$ increases slightly with the temperature. And there is no qualitative change at different values of the critical temperature.

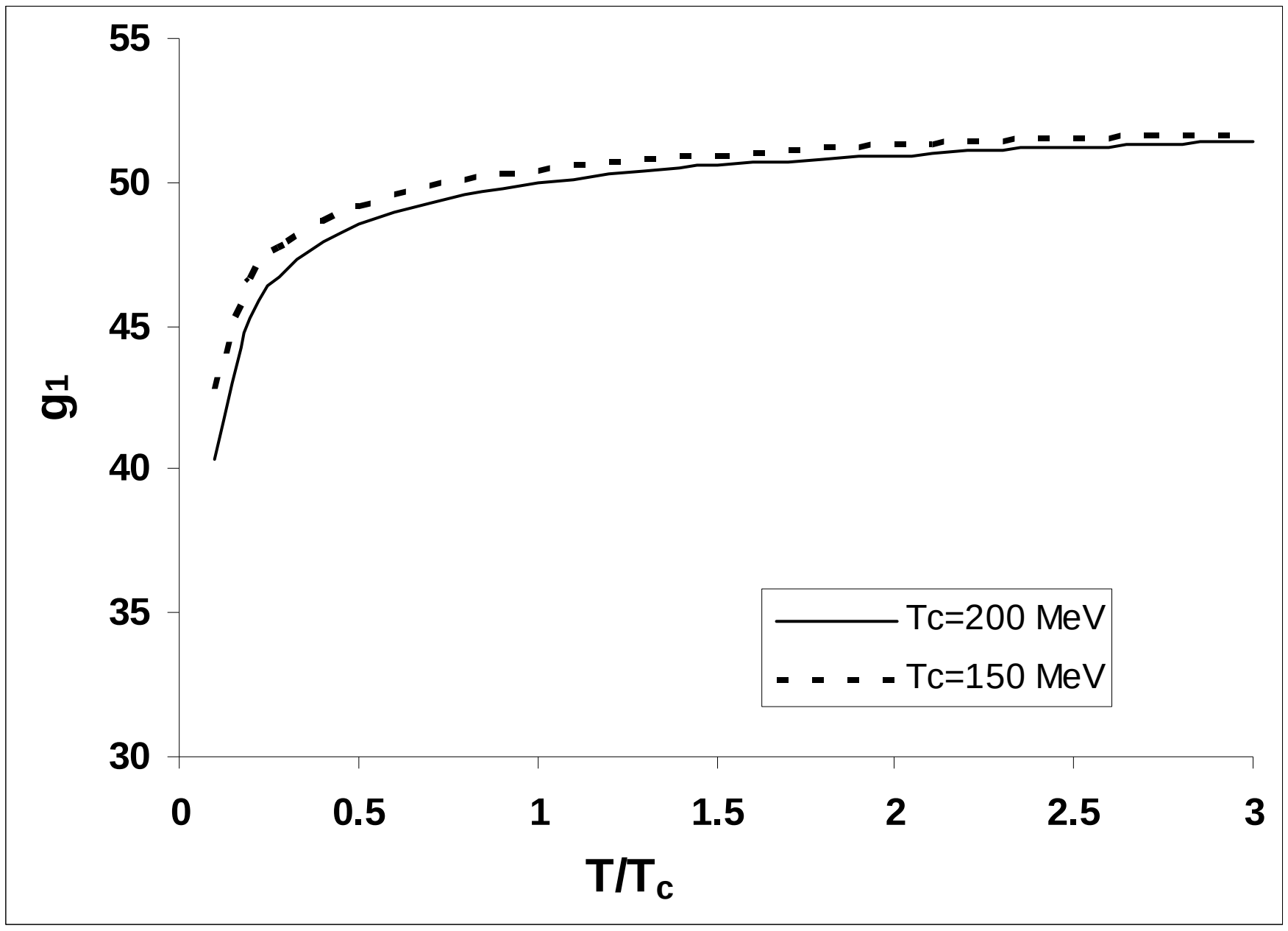

**Fig.(2) The degree of freedom in the deconfined QGP phase $g_1$ versus $T/T_c$.**

Fig. (3) shows the difference between $g_1$ and $g_2$ ($\Delta g$) versus $T/T_c$ in which $g_1$ is the effective one. Because of its dependence on the running coupling constant whereas $g_2$ has constant value g2=12.25. So far we conclude that g1 is the effective quantity in these calculations through the switch the strong coupling constant to the dependent temperature running coupling constant

Fig. (4) shows the plotting between the latent heat $B/T_C^4$ versus $T/T_c$ at different values of the critical temperatures $T_c$ = 150, 200 MeV. The solid curve is the calculation of $B/T_C^4$ at $T_c$ =200 MeV, and the dashed curve is the same calculation at $T_c$= 150 MeV. It is obvious that, at $T = T_c$ the value of $B/T_C^4 \approx 3.2$.

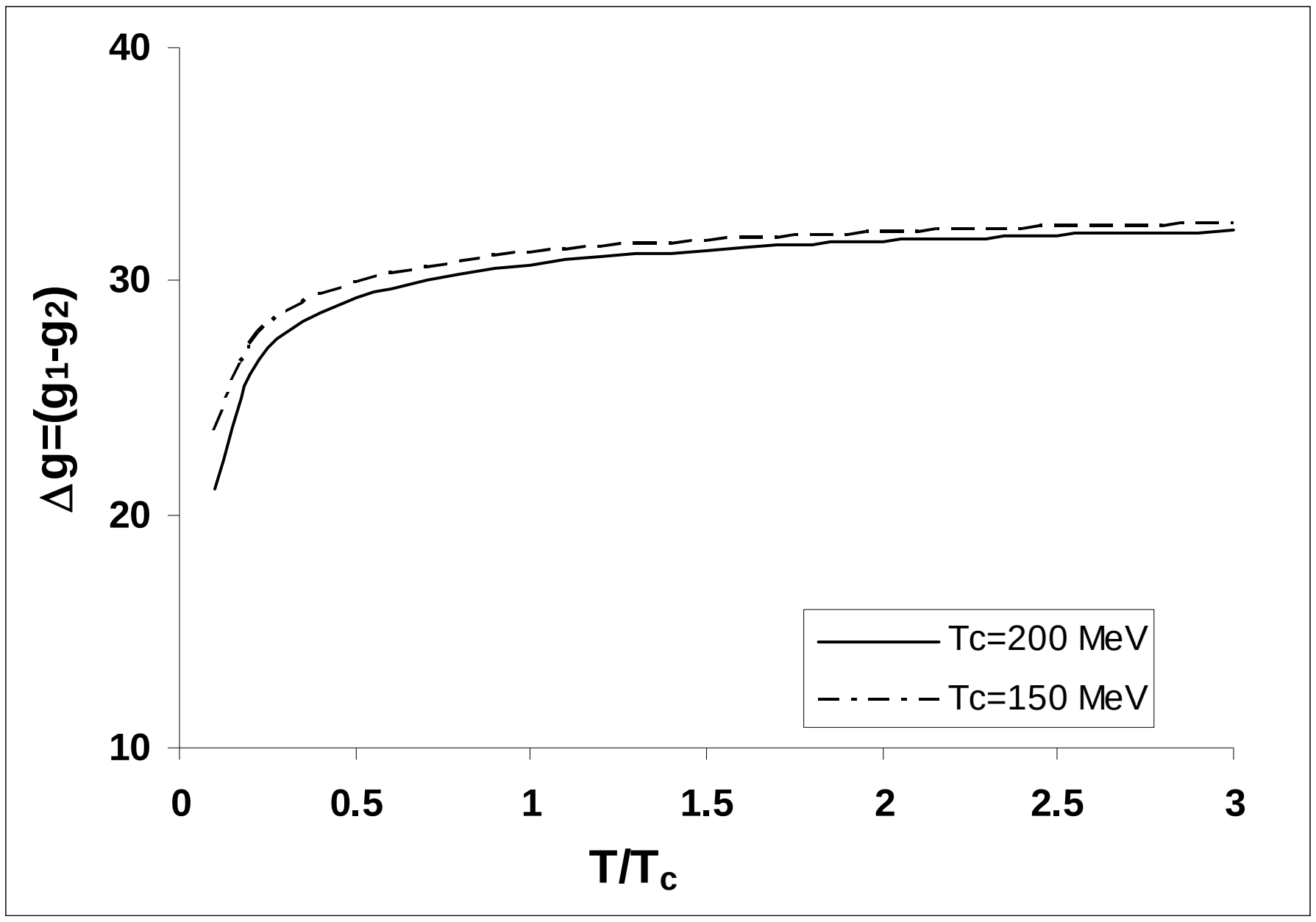

**Fig.(3) The difference between the deconfined and confined degrees of freedom $\Delta g$ versus T/T$_c$.**

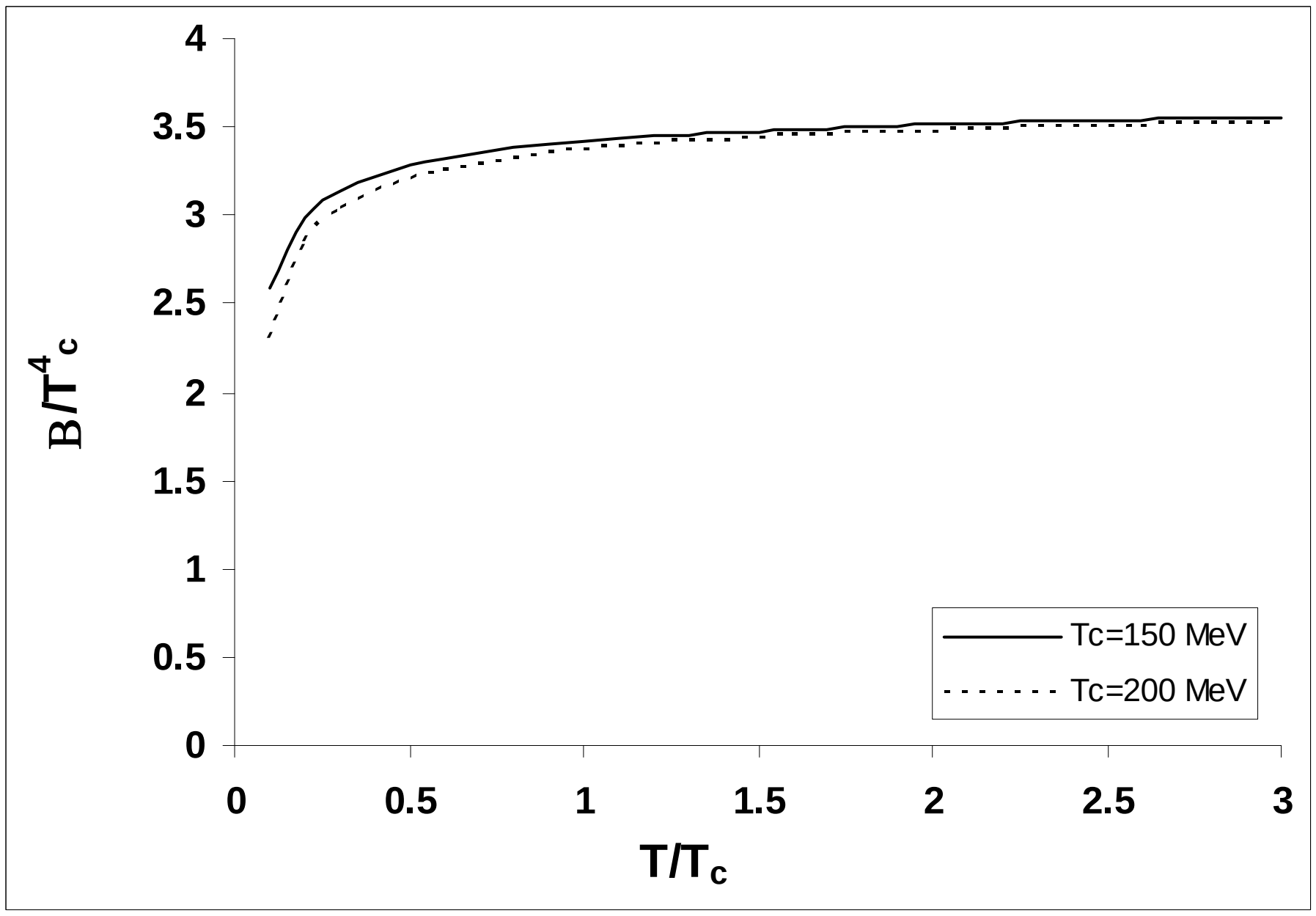

**Fig.(4) The latent heat between the confined and deconfined phase transition versus T/T$_c$.**

Fig. (5) is the most important curve according to the approach we have used in this paper by inserting the running coupling constant $\alpha_s(T)$ instead of the strong coupling constant $\alpha_s$. This figure shows the relation between the critical pressure $P_c / T_c^4$ versus T/T$_c$ at critical temperature T$_c$= 200 MeV. From this figure one can determine the critical pressure at T=T$_c$, then at T=T$_c$ the value of $P_c / T_c^4 \approx 2.07$.

Then one can calculate such curve at different values of T$_c$, and determine the different critical values of the pressure. At this time one has ranges of the critical pressure values according to the ranges of the critical temperatures. This allows a wide range of the temperature of the studying. Also from the computed curves at different critical temperature one can expect the range which is convenient before deconfinment and the range after deconfinment from the behavior of the different curves, then one has also a range from the phase transition point.

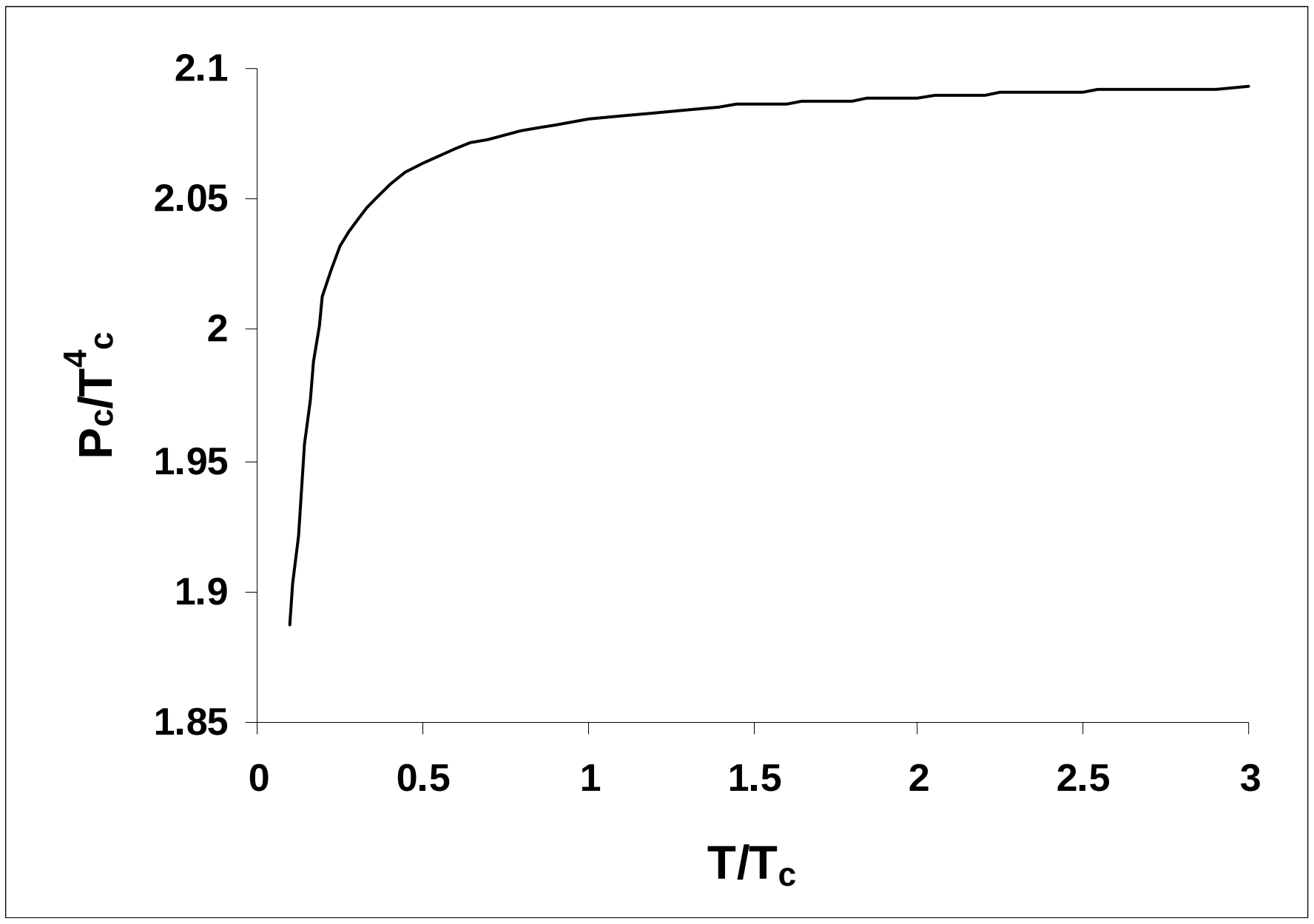

**Fig. (5) The critical pressure $P_c / T_c^4$ versus T/T$_c$, T$_c$=200 MeV.**

## Conclusion

In this work we have studied the effect of the running coupling constant instead of the strong coupling constant to calculate the degree of freedom in the deconfined QGP phase of the early Universe, the latent heat and the critical pressure. We have concluded that, by inserting the dependent temperature running coupling in the calculations allows to get a range for the critical pressure at different T=T$_c$.